\documentclass[english,draftcls, onecolumn]{IEEEtran}
\usepackage[T1]{fontenc}
\usepackage[latin9]{inputenc}
\usepackage{array}
\usepackage{float}
\usepackage{amsthm}
\usepackage{amsmath}
\usepackage{amssymb}
\usepackage{graphicx}

\makeatletter

\providecommand{\tabularnewline}{\\}
\floatstyle{ruled}
\newfloat{algorithm}{tbp}{loa}
\providecommand{\algorithmname}{Algorithm}
\floatname{algorithm}{\protect\algorithmname}

 \usepackage{algolyx}
  \theoremstyle{plain}
  \newtheorem{lem}{\protect\lemmaname}
  \theoremstyle{plain}
  \newtheorem{thm}{\protect\theoremname}

\usepackage{cite}
\allowdisplaybreaks
\IEEEoverridecommandlockouts
\@ifundefined{showcaptionsetup}{}{%
 \PassOptionsToPackage{caption=false}{subfig}}
\usepackage{subfig}
\makeatother

\usepackage{babel}
\providecommand{\lemmaname}{Lemma}
\providecommand{\theoremname}{Theorem}

\begin{document}

\title{Concurrent learning for parameter estimation using dynamic state-derivative
estimators%
\thanks{Rushikesh Kamalapurkar and Warren E. Dixon are with the Department
of Mechanical and Aerospace Engineering, University of Florida, Gainesville,
FL, USA. Email: \{rkamalapurkar, wdixon\}@ufl.edu. Benjamin Reish
and Girish Chowdhary are with the Department of Mechanical and Aerospace
Engineering, Oklahoma State University, Stillwater, OK, USA. Email:
reish@ostatemail.okstate.edu and girish.chowdhary@okstate.edu%
}}

\author{Rushikesh Kamalapurkar, Ben Reish, Girish Chowdhary, and Warren E.
Dixon}
\maketitle
\begin{abstract}
A concurrent learning (CL)-based parameter estimator is developed
to identify the unknown parameters in a linearly parameterized uncertain
control-affine nonlinear system. Unlike state-of-the-art CL techniques
that assume knowledge of the state-derivative or rely on numerical
smoothing, CL is implemented using a dynamic  state-derivative estimator.
A novel purging algorithm is introduced to discard possibly erroneous
data recorded during the transient phase for concurrent learning.
Since purging results in a discontinuous parameter adaptation law,
the closed-loop error system is modeled as a switched system. Asymptotic
convergence of the error states to the origin is established under
a \textit{persistent} excitation condition, and the error states are
shown to be ultimately bounded under a \textit{finite} excitation
condition. Simulation results are provided to demonstrate the effectiveness
of the developed parameter estimator.
\end{abstract}

\section{Introduction}

Modeling and identification of input-output relationships of nonlinear
dynamical systems has been a long-standing active area of research.
A variety of offline techniques have been developed for system identification;
however, when models are used for online feedback control, the ability
to adapt to changing environment and the ability to learn from input-output
data are desirable. Motivated by applications in feedback control,
online system identification techniques are investigated in results
such as \cite{Sureshbabu.Farrell1999,Nelles2001,Jaeger2002,Angelov.Filev2004}
and the references therein. 

Parametric methods such as linear parameterization, neural networks
and fuzzy logic systems approximate the system identification problem
by a finite-dimensional parameter estimation problem, and hence, are
popular tools for online nonlinear system identification. Parametric
models have been widely employed for adaptive control of nonlinear
systems. In general, adaptive control methods do not require or guarantee
convergence of the parameters to their true values. However, parameter
convergence has been shown to improve robustness and transient performance
of adaptive controllers (cf. \cite{Duarte.Narendra1989,Krstic.Kokotovic.ea1993,Lavretsky2009,Chowdhary.Johnson2011a}).
Parametric models have also been employed in optimal control techniques
such as model-based predictive control (MPC) (cf. \cite{Fukushima.Kim.ea2007,Adetola.DeHaan.ea2009,Chowdhary.Muehlegg.ea2013,Aswani.Gonzalez.ea2013})
and model-based reinforcement learning (MBRL) (cf. \cite{Abbeel.Quigley.ea2006,Mitrovic.Klanke.ea2010,Deisenroth.Rasmussen2011,Kamalapurkar.Walters.ea2013}).
In MPC and MBRL, the controller is developed based on the parameter
estimates; hence, stability of the closed-loop system and the performance
of the developed controller critically depend on convergence of the
parameter estimates to their ideal values.

Data-driven concurrent learning (CL) techniques are developed in results
such as \cite{Chowdhary2010a,Chowdhary.Johnson2011a,Chowdhary.Yucelen.ea2012},
where recorded data is concurrently used with online data to achieve
parameter convergence under a relaxed \textit{finite} excitation condition
as opposed to the \textit{persistent} excitation (PE) condition required
by traditional adaptive control methods. CL techniques employ the
fact that a direct formulation of the parameter estimation error can
be obtained provided the state-derivative is known or its estimate
is otherwise available through techniques such as fixed-point smoothing
\cite{Muehlegg.Chowdhary.ea2012}. The parameter estimation error
can then be used in a gradient-based adaptation algorithm to drive
the parameter estimates to their ideal values. If exact derivatives
are not available, the parameter estimation error can be shown to
decay to a neighborhood of the origin provided accurate estimates
of the state-derivatives are available, where the size of the neighborhood
depends on the derivative estimation error \cite{Muehlegg.Chowdhary.ea2012}.
Experimental results such as \cite{Chowdhary.Johnson2011a} demonstrate
that, since derivatives at past data points are required, noncausal
numerical smoothing techniques can be used to generate satisfactory
estimates of state-derivatives. Under Gaussian noise, smoothing is
guaranteed to result in the best possible linear estimate corresponding
to the available data \cite[Section 5.3]{Gelb1974}; however, in general,
the derivative estimation error resulting from numerical smoothing
can not be quantified a priori. Furthermore, numerical smoothing requires
additional processing and storage of data over a time-window that
contains the point of interest. Hence, the problem of achieving parameter
convergence under relaxed excitation conditions without using numerical
differentiation is also motivated.

In this paper, an observer is employed to estimate the state-derivative.
The derivative estimate generated by the observer converges exponentially
to a neighborhood of the actual state-derivative. However, in the
transient phase, the derivative estimation errors can be large. Since
CL relies on repeated use recorded data, large transient errors present
a challenge in the development of a CL-based parameter estimator.
If the derivative estimation errors at the points recorded in the
history stack are large, then the corresponding errors in the parameter
estimates will be large. Motivated by the results in \cite{Reish.Chowdhary2014},
the aforementioned challenge is addressed in this paper by designing
a novel purging algorithm to purge possibly erroneous data from the
history stack. The closed-loop error system along with the purging
algorithm is modeled as a switched nonlinear dynamical system. Provided
enough data can be recorded to populate the history stack after each
purge, the developed method ensures asymptotic convergence of the
error states to the origin. 

The PE condition can be shown to be sufficient to ensure that enough
data can be recorded to populate the history stack after each purge.
Since PE can be an impractical requirement in many applications, this
paper examines the behavior of the switched error system under a relaxed
finite excitation condition. Specifically, provided the system states
are exciting over a sufficiently long finite time-interval, the error
states decay to an ultimate bound. Furthermore, the ultimate bound
can be made arbitrarily small by increasing the learning gains. Simulation
results are provided to demonstrate the effectiveness of the developed
method under measurement noise.

\section{System dynamics}

The system dynamics are assumed to be nonlinear, uncertain, and control-affine$^{\ref{fn:CLNXDThe-focus-of}}$,
described by the differential equation
\begin{equation}
\dot{x}=f\left(x,u\right),\label{eq:CLNoXDotDyn}
\end{equation}
where the function $f:\mathbb{R}^{n\times m}\to\mathbb{R}^{n}$ is
locally Lipschitz. It is assumed that the dynamics can be split into
a known component and an uncertain component with parametric uncertainties
that are linear in the parameters. That is, $f\left(x,u\right)=f_{1}\left(x\right)+g\left(x\right)u+Y\left(x\right)\theta$,
where $f_{1}:\mathbb{R}^{n}\to\mathbb{R}^{n}$, $g:\mathbb{R}^{n}\to\mathbb{R}^{n\times m}$
and $Y:\mathbb{R}^{n}\to\mathbb{R}^{n\times P}$ are known and locally
Lipschitz continuous. The constant parameter vector $\theta\in\mathbb{R}^{P}$
is unknown, with a known bound $\overline{\theta}\in\mathbb{R}_{>0}$
such that $\left\Vert \theta\right\Vert <\overline{\theta}$. The
objective is to design a parameter estimator to estimate the unknown
parameters. The system input is assumed to be a stabilizing controller
such that $x,\dot{x},u\in\mathcal{L}_{\infty}.$%
\footnote{\label{fn:CLNXDThe-focus-of}The focus of this paper is adaptive estimation,
and not control design. Even though most adaptive controllers are
designed based on an estimate of the unknown parameters, parameter
estimation can often be decoupled from control design. For example,
the adaptive controller in \cite{Patre2011} can guarantee $x,\dot{x},u\in\mathcal{L}_{\infty}$
for a wide class of adaptive update laws. Under the additional assumption
that $\dot{u}\in\mathcal{L}_{\infty}$, the developed technique can
be extended to include linearly parameterized nonaffine systems, that
is, $f\left(x,u\right)=f_{1}\left(x\right)+g\left(x\right)u+Y\left(x,u\right)\theta$.%
} The system state $x$ is assumed to be available for feedback, and
the state-derivative $\dot{x}$ is assumed to be unknown.

\section{CL-based adaptive derivative estimation}

Let $\hat{x}\in\mathbb{R}^{n}$ and $\dot{\hat{x}}\in\mathbb{R}^{n}$
denote estimates of the state $x$, and the state-derivative $\dot{x}$,
respectively. Let $\hat{\theta}\in\mathbb{R}^{P}$ denote an estimate
of the unknown vector $\theta$. To achieve convergence of the estimate
$\hat{\theta}$ to the ideal parameter vector $\theta,$ a CL-based
parameter estimator is designed. The motivation behind CL is to adjust
the parameter estimates based on an estimate of the parameter estimation
error $\tilde{\theta},$ defined as $\tilde{\theta}\triangleq\theta-\hat{\theta}$,
in addition to the state estimation error $\tilde{x}.$ Since $\tilde{\theta}$
is not directly measurable, the subsequent development exploits the
fact that the term $Y\left(x\right)\tilde{\theta}$ can be computed
as $Y\left(x\right)\tilde{\theta}=\dot{x}-f_{1}\left(x\right)-g\left(x\right)u-Y\left(x\right)\hat{\theta}$,
provided measurements of the state-derivative $\dot{x}$ are available.
In CL results such as \cite{Chowdhary2010a,Chowdhary.Johnson2011a,Chowdhary.Yucelen.ea2012,Kingravi.Chowdhary.ea2012},
it is assumed that the state-derivatives can be computed with sufficient
accuracy at a past time instance by numerically differentiating the
recorded data. An approximation of the parameters estimation errors
is then computed as $Y\left(x_{j}\right)\tilde{\theta}+\overline{d}=\dot{\overline{x}}_{j}-f_{1}\left(x_{j}\right)-g\left(x_{j}\right)u_{j}-Y\left(x_{j}\right)\hat{\theta}$,
where $x_{j}$ denotes the system state at a past time instance $t_{j}$,
$\dot{\overline{x}}_{j}$ denotes the numerically computed state-derivative
at $t_{j}$, and $\overline{d}$ is a constant of the order of the
error between $\dot{x}_{j}$ and $\dot{\overline{x}}_{j}$. While
the results in \cite{Muehlegg.Chowdhary.ea2012} establish that, provided
$\overline{d}$ is bounded, the parameter estimation error $\tilde{\theta}$
can be shown to decay to a ball around the origin, the focus is on
the analysis of the effects of the differentiation error, and not
on development of algorithms to reduce the parameter estimation error.

In this paper, a dynamically generated estimate of the state-derivative
is used instead of numerical smoothing. The parameter estimation error
is computed at a past recorded data point as $Y\left(x_{j}\right)\tilde{\theta}-\dot{\tilde{x}}_{j}=\dot{\hat{x}}_{j}-f_{1}\left(x_{j}\right)-g\left(x_{j}\right)u_{j}-Y\left(x_{j}\right)\hat{\theta}$,
where $\dot{\tilde{x}}_{j}\triangleq\dot{x}_{j}-\dot{\hat{x}}_{j}$.
To facilitate the design, let $\mathcal{H}\triangleq\left\{ \left(\dot{\hat{x}}_{j},x_{j},u_{j}\right)\right\} _{j=1}^{M}$
be a history stack containing recorded values of the state, the control,
and the state-derivative estimate. Each tuple $\left(\dot{\hat{x}}_{j},x_{j},u_{j}\right)$
is referred to as a data-point in $\mathcal{H}$. A history stack
$\mathcal{H}$ is called ``full rank'' if the state vectors recorded
in $\mathcal{H}$ satisfy $\mbox{rank}\left(\sum_{j=1}^{M}Y^{T}\left(x_{j}\right)Y\left(x_{j}\right)\right)=P$.
Based on the subsequent Lyapunov-based stability analysis, the history
stack is used to update the estimate $\hat{\theta}$ using the following
update law:
\begin{equation}
\dot{\hat{\theta}}\!=\! k\Gamma\!\!\sum_{j=1}^{M}\!\! Y^{T}\!\left(\! x_{j}\!\right)\!\left(\!\dot{\hat{x}}_{j}\!-\! f_{1}\!\left(\! x_{j}\!\right)\!-\! g\!\left(\! x_{j}\!\right)\! u_{j}\!-\! Y\!\left(\! x_{j}\!\right)\!\hat{\theta}\!\right)\!+\!\Gamma Y^{T}\!\left(\! x\!\right)\!\tilde{x},\label{eq:CLNoXDotThetaHat}
\end{equation}
where $\tilde{x}=x-\hat{x}\in\mathbb{R}^{n}$, $\Gamma\in\mathbb{R}^{P\times P}$
and $k\in\mathbb{R}$ are positive and constant learning gains. 

The update law in (\ref{eq:CLNoXDotThetaHat}) drives the parameter
estimation error to a ball around the origin, the size of which is
of the order of $\dot{\tilde{x}}_{j}.$ Hence, to achieve a lower
parameter estimation error, it is desirable to drive $\dot{\tilde{x}}$
to the origin. Based on the subsequent Lyapunov-based stability analysis,
the following adaptive  estimator is designed to generate the state-derivative
estimates:
\begin{align}
\dot{\hat{x}} & =\gamma_{1}Y\left(x\right)\hat{\theta}+f_{1}\left(x\right)+g\left(x\right)u+\left(k_{1}+\alpha_{1}\right)\tilde{x}+\mu,\nonumber \\
\dot{\mu} & =\left(k_{1}\alpha_{1}+1\right)\tilde{x},\label{eq:CLNoXDotEst}
\end{align}
where $\mu\in\mathbb{R}^{n}$ is an auxiliary signal and $k_{1},\alpha_{1}\in\mathbb{R}_{>0}$
and $\gamma_{1}\in\left[0,1\right]$ are positive constant learning
gains.

\section{Algorithm to record the history stack}

\subsection{Purging of history stacks}

The state-derivative estimator in (\ref{eq:CLNoXDotEst}) relies on
 feedback of the state estimation error $\tilde{x}.$ In general,
 feedback results in large transient estimation errors. Hence, the
state-derivative estimation errors $\dot{\tilde{x}}_{j}$ associated
with the tuples $\left(\dot{\hat{x}}_{j},x_{j},u_{j}\right)$ recorded
in the transient phase can be large. The results in \cite{Muehlegg.Chowdhary.ea2012}
imply that the parameter estimation errors can be of the order of
$\max_{j}\left\Vert \dot{\tilde{x}}_{j}\right\Vert .$ Hence, if a
history stack containing data-points with large derivative estimation
errors is used for CL, then the parameter estimates will converge
but the resulting parameter estimation errors can be large. To mitigate
the aforementioned problem, this paper introduces a new algorithm
that purges the erroneous data in the history stack as soon as more
data is available. Since the  estimator in (\ref{eq:CLNoXDotEst})
results in exponential convergence of $\dot{\hat{x}}$ to a neighborhood
of $\dot{x},$ newer data is guaranteed to represent the system better
than older data, resulting in a lower steady-state parameter estimation
error. The following section details the proposed algorithm.

\subsection{Algorithm to record the history stack}

The history stack $\mathcal{H}$ is initialized arbitrarily to be
full rank. An arbitrary full rank initialization of $\mathcal{H}$
results in a $\sigma-$modification (cf. \cite{Ioannou.Kokotovic1983})
like adaptive update law that keeps the parameter estimation errors
bounded. However, since the history stack may contain erroneous data,
the parameters may not converge to their ideal values. 

In the following, a novel algorithm is developed to keep the history
stack current by purging the existing (and possibly erroneous) data
and replacing it with current data. The data collected from the system
is recorded in an auxiliary history stack $\mathcal{G}\triangleq\left\{ \left(\dot{\hat{x}}_{j}^{G},x_{j}^{G},u_{j}^{G}\right)\right\} _{j=1}^{M}$.
The history stack $\mathcal{G}$ is initialized such that $\left(\dot{\hat{x}}_{j}^{G}\left(0\right),x_{j}^{G}\left(0\right),u_{j}^{G}\left(0\right)\right)=\left(\mathbf{0}_{n},\mathbf{0}_{n},\mathbf{0}_{m}\right)$
and is populated using a singular value maximization algorithm \cite{Chowdhary2010a}.
Once the history stack $\mathcal{G}$ becomes full rank with a minimum
singular value that is above a (static or dynamic) threshold, $\mathcal{H}$
is replaced with $\mathcal{G}$, and $\mathcal{G}$ is purged.%
\footnote{Techniques such as probabilistic confidence checks (cf. \cite{Reish.Chowdhary2014})
can also be utilized to initiate purging. The following analysis is
agnostic with respect to the trigger used for purging provided $\mathcal{G}$
is full rank at the time of purging and the dwell time $\mathcal{T}$
is maintained between two successive purges.%
} In this paper, a dynamic threshold is used which is set to be a fraction
of the highest encountered minimum singular value corresponding to
$\mathcal{H}$ up to the current time.

In the subsequent Algorithm \ref{alg:CLNoXDotpurgeDwell}, a piece-wise
constant function $\delta:\mathbb{R}_{\geq0}\to\mathbb{R}_{\geq0}$,
initialized to zero, stores the last time instance when $\mathcal{H}$
was updated and a piecewise constant function $\eta:\mathbb{R}_{\geq0}\to\mathbb{R}_{\geq0}$
stores the highest encountered value of $\sigma_{\mbox{min}}\left(\sum_{j=1}^{M}Y^{T}\left(x_{j}\right)Y\left(x_{j}\right)\right)$
up to time $t$, where $\sigma_{\mbox{min}}$ denotes the minimum
singular value. The constant $\xi\in\left(0,1\right)$ denotes the
threshold fraction used to purge the history stack, and $\mathcal{T}\in\mathbb{R}$
is an adjustable positive constant.
\begin{algorithm}
\begin{algor}
\item [{if}] a data point is available

\begin{algor}
\item [{if}] $\mathcal{G}$ is not full

\begin{algor}
\item [{do}] add the data to $\mathcal{G}$ 
\end{algor}
\item [{else}]~

\begin{algor}
\item [{do}] add the data to $\mathcal{G}$ if $\sigma_{\mbox{min}}\left(\sum_{j=1}^{M}Y^{T}\left(x_{j}^{G}\right)Y\left(x_{j}^{G}\right)\right)$
increases
\end{algor}
\item [{endif}]~
\item [{if}] $\sigma_{\mbox{min}}\left(\sum_{j=1}^{M}Y^{T}\left(x_{j}^{G}\right)Y\left(x_{j}^{G}\right)\right)\geq\xi\eta\left(t\right)$

\begin{algor}
\item [{if}] $t-\delta\left(t\right)\geq\mathcal{T}$ 

\begin{algor}
\item [{do}] $\mathcal{H}\leftarrow\mathcal{G}$ and $\mathcal{G}\leftarrow0$
(purge $\mathcal{G}$)
\item [{do}] $\delta\left(t\right)\leftarrow t$
\item [{if}] $\eta\left(t\right)<\sigma_{\mbox{min}}\left(\sum_{j=1}^{M}Y^{T}\left(x_{j}\right)Y\left(x_{j}\right)\right)$

\begin{algor}
\item [{do}] $\eta\left(t\right)\leftarrow\sigma_{\mbox{min}}\left(\sum_{j=1}^{M}Y^{T}\left(x_{j}\right)Y\left(x_{j}\right)\right)$
\end{algor}
\item [{endif}]~
\end{algor}
\item [{endif}]~
\end{algor}
\item [{endif}]~
\end{algor}
\item [{endif}]~
\end{algor}
\protect\caption{\label{alg:CLNoXDotpurgeDwell}History stack purging with dwell time}
\end{algorithm}

The following analysis establishes that if the system states are \textit{persistently}
exciting (in a sense that will be made clear in Theorem \ref{thm:CLNoXDotAsymptotic})
then the parameter estimation error asymptotically decays to zero.
Furthermore, it is also established that if the system states are
exciting over a \textit{finite} period of time, then the parameter
estimation error can be made as small as desired provided $\mathcal{T}$
and the learning gains are selected based on the sufficient conditions
introduced in Theorem \ref{thm:CLNoXDotUUB}.

\section{Analysis}

\subsection{\label{sub:CLNoXDotAsymptotic-convergence-with}Asymptotic convergence
with persistent excitation}

Purging of the history stack $\mathcal{H}$ implies that the resulting
closed-loop system is a switched system, where each subsystem corresponds
to a history stack, and each purge indicates a switching event.%
\footnote{Since a switching event in Algorithm \ref{alg:CLNoXDotpurgeDwell}
occurs only when the auxiliary history stack is full, Zeno behavior
is avoided by design.%
} To facilitate the analysis, let $\rho:\mathbb{R}_{\geq0}\to\mathbb{N}$
denote a switching signal such that $\rho\left(0\right)=1$, and $\rho\left(t\right)=j+1$,
where $j$ denotes the number of times the update $\mathcal{H}\leftarrow\mathcal{G}$
was carried out over the time interval $\left(0,t\right)$. In the
following, the subscript $s\in\mathbb{N}$ denotes the switching index,
and $\mathcal{H}_{s}$ denotes the history stack corresponding to
the $s$\textsuperscript{th} subsystem (i.e., the history stack active
during the time interval $\left\{ t\mid\rho\left(t\right)=s\right\} $),
containing the elements $\left\{ \left(\dot{\hat{x}}_{sj},x_{sj},u_{sj}\right)\right\} _{j=1}^{M}$.
To simplify the notation, let 
\begin{equation}
A_{s}=\sum_{j=1}^{M}Y^{T}\left(x_{sj}\right)Y\left(x_{sj}\right),\: Q_{s}=\sum_{j=1}^{M}Y^{T}\left(x_{sj}\right)\dot{\tilde{x}}_{sj}.\label{eq:CLTNoXDotPQ}
\end{equation}
Note that $A_{s}:\mathbb{R}_{\geq0}\to\mathbb{R}^{P\times P}$ and
$Q_{s}:\mathbb{R}_{\geq0}\to\mathbb{R}^{P\times1}$ are piece-wise
constant functions of time. For ease of exposition, the constant $\mathcal{T}$
introduced in Algorithm \ref{alg:CLNoXDotpurgeDwell} is set to zero
for the case where persistent excitation is available.

Algorithm \ref{alg:CLNoXDotpurgeDwell} ensures that there exists
a constant $\underline{a}>0$ such that $\lambda_{\min}\left\{ A_{s}\right\} \geq\underline{a},\:\forall s\in\mathbb{N},$
where $\lambda_{\mbox{min}}$ denotes the minimum eigenvalue. Since
the state $x$ remains bounded by assumption, there exists a constant
$\overline{A}$ such that $\left\Vert A_{s}\right\Vert \leq\overline{A},\:\forall s\in\mathbb{N}.$

Using (\ref{eq:CLNoXDotThetaHat}) and (\ref{eq:CLTNoXDotPQ}), the
dynamics of the parameter estimation error $\tilde{\theta}$ can be
written as 
\begin{equation}
\dot{\tilde{\theta}}=-\Gamma Y^{T}\left(x\right)\tilde{x}-k\Gamma A_{s}\tilde{\theta}+k\Gamma Q_{s}.\label{eq:CLNoXDotThetaTilde}
\end{equation}
To establish convergence of the state-derivative estimates, a filtered
tracking error $r\in\mathbb{R}^{n}$ is defined as $r\triangleq\dot{\tilde{x}}+\alpha_{1}\tilde{x}.$
Using (\ref{eq:CLNoXDotDyn}), (\ref{eq:CLNoXDotEst}), and (\ref{eq:CLNoXDotThetaTilde}),
the time derivative of the filtered tracking error can be expressed
as, 
\begin{equation}
\dot{r}=F\left(x,u\right)\tilde{\theta}-k\gamma_{1}Y\left(x\right)\Gamma A_{s}\tilde{\theta}-\gamma_{1}Y\left(x\right)\Gamma Y^{T}\left(x\right)\tilde{x}-\tilde{x}+k\gamma_{1}Y\left(x\right)\Gamma Q_{s}-k_{1}r,\label{eq:CLNoXDotrdot}
\end{equation}
where $F\left(x,u\right)\triangleq\gamma_{1}\nabla Y\left(x\right)Y\left(x\right)\theta+\gamma_{1}\nabla Y\left(x\right)f_{1}\left(x\right)+\gamma_{1}\nabla Y\left(x\right)g\left(x\right)u$
and $\nabla Y\left(x\right)\triangleq\partial Y\left(x\right)/x$.

To facilitate the stability analysis, let $\overline{F},$ $\overline{F}_{1},$
$\overline{x},$ $\overline{Y}$, and $\overline{\Gamma}$ be constants
such that 
\begin{gather}
\left\Vert F\left(x\left(t\right),u\left(t\right)\right)\right\Vert \leq\overline{F},\:\left\Vert Y\left(x\left(t\right)\right)\right\Vert \leq\overline{Y},\:\left\Vert \Gamma\right\Vert =\overline{\Gamma},\nonumber \\
\left\Vert f_{1}\left(x\left(t\right)\right)+Y\left(x\left(t\right)\right)\theta+g\left(x\left(t\right)\right)u\left(t\right)\right\Vert \leq\overline{F}_{1},\:\left\Vert x\left(t\right)\right\Vert \leq\overline{x},\label{eq:CLNoXDotBounds}
\end{gather}
for all $t\in\mathbb{R}_{\geq0}$. The following stability analysis
is split into three parts. Under the temporary assumptions that the
error states $\tilde{\theta}$, $\tilde{x}$, and $r$ are bounded
at a switching instance and that the norms of the state-derivative
estimates stored in the history stack are bounded, it is established
in Part 1 that the error states $\tilde{\theta}$, $\tilde{x}$, and
$r$ decay to a bound before the next switching instance, where the
bound depends on the derivative estimation errors. Under the temporary
assumption that the error states $\tilde{x}$ and $r$ are bounded
at a switching instance, it is established in Part 2 that the derivative
estimation error $\dot{\tilde{x}}$ can be made arbitrarily small
before the next switching instance by increasing the learning gains.
In Part 3, the temporary assumptions in Parts 1 and 2 are relaxed
through an inductive argument where the results from Part 1 and Part
2 are used to conclude asymptotic convergence of the error states
$\tilde{\theta},$ $\tilde{x}$, and $r$ to the origin.

\subsubsection*{Part 1: Boundedness of the error signals}

Let $Z\triangleq\left[\begin{array}{ccc}
r^{T} & \tilde{x}^{T} & \tilde{\theta}^{T}\end{array}\right]^{T}\in\mathbb{R}^{2n+L}$ and let $V:\mathbb{R}^{2n+P}\to\mathbb{R}_{\geq0}$ denote a candidate
Lyapunov function defined as 
\begin{align}
V\left(Z\right) & \triangleq\frac{1}{2}r^{T}r+\frac{1}{2}\tilde{x}^{T}\tilde{x}+\frac{1}{2}\tilde{\theta}^{T}\Gamma^{-1}\tilde{\theta},\label{eq:CLNoXDotV}
\end{align}
Using the Raleigh-Ritz Theorem, the Lyapunov function $V$ can be
bounded as
\begin{equation}
\underline{v}\left\Vert Z\right\Vert ^{2}\leq V\left(Z\right)\leq\overline{v}\left\Vert Z\right\Vert ^{2},\label{eq:CLNoXDotVBounds}
\end{equation}
where $\overline{v}\triangleq\frac{1}{2}\max\left\{ 1,\lambda_{\max}\left\{ \Gamma^{-1}\right\} \right\} ,$
$\underline{v}\triangleq\frac{1}{2}\min\left\{ 1,\lambda_{\min}\left\{ \Gamma^{-1}\right\} \right\} .$
The subsequent stability analysis assumes that the learning gains
$k,$ $k_{1}$, and $\alpha_{1},$ and the matrices $A_{s}$ satisfy
the following sufficient gain conditions:%
\footnote{The sufficient conditions can be satisfied provided the gains $k_{1}$
and $\alpha_{1}$ are selected large enough.%
}
\begin{equation}
\underline{a}>\frac{3\overline{Y}^{2}}{k\alpha_{1}}+\frac{4\overline{F}^{2}}{kk_{1}}+\frac{4k\overline{Y}^{2}\overline{\Gamma}^{2}\overline{A}^{2}}{k_{1}},\quad k_{1}>\frac{6\overline{Y}^{4}\overline{\Gamma}^{2}}{\alpha_{1}}.\label{eq:CLTNoXDotL1Suff}
\end{equation}
The following Lemma establishes boundedness of the error state $Z$.
\begin{lem}
\label{lem:CLTNoXDotThTBound}Let $T\in\mathbb{R}_{>0}$ be a constant
such that $\rho\left(\tau\right)=s$, for all $\tau\in\left[t,t+T\right)$.
Assume temporarily that there exist constants $\overline{H}_{s},\overline{V}_{s}\in\mathbb{R}_{>0}$
such that the elements of $\mathcal{H}_{s}$ satisfy $\left\Vert \dot{\hat{x}}_{j}\right\Vert \leq\overline{H}_{s}$,
for all $j\in\left\{ 1,\cdots M\right\} $, and that the candidate
Lyapunov function satisfies $V\left(Z\left(t\right)\right)\leq\overline{V}_{s}$.
Then, the candidate Lyapunov function $V$ is bounded as
\begin{equation}
V\!\left(Z\!\left(\tau\right)\right)\!\leq\!\left(\!\overline{V}_{s}\!-\!\frac{\overline{v}}{v}\iota_{s}\!\right)\! e^{-\frac{v}{\overline{v}}\left(\tau-t\right)}\!+\!\frac{\overline{v}}{v}\iota_{s},\:\forall\tau\!\in\!\left[t,t+T\right),\label{eq:CLNoXDotVDecay}
\end{equation}
where $\iota_{s}\triangleq\left(\frac{k}{2\underline{a}}+\frac{k^{2}\overline{Y}^{2}\overline{\Gamma}^{2}}{k_{1}}\right)\overline{Q}_{s}^{2}.$
Furthermore, the parameter estimation error can be bounded as 
\begin{equation}
\left\Vert \tilde{\theta}\left(\tau\right)\right\Vert \leq\theta_{s},\:\forall\tau\in\left[t,t+T\right)\label{eq:CLNoXDotThetaTildeBound}
\end{equation}
where $\theta_{s}\triangleq\sqrt{\frac{\:\overline{v}\:}{\underline{v}}}\max\left\{ \sqrt{\overline{V}_{s}},\sqrt{\iota_{s}}\right\} ,$
$v\triangleq\min\left\{ \frac{k\underline{c}}{4},\frac{\alpha_{1}}{3},\frac{k_{1}}{8}\right\} ,$
and $\overline{Q}_{s}\triangleq\left\Vert Q_{s}\right\Vert $.\end{lem}
\begin{IEEEproof}
Using (\ref{eq:CLNoXDotThetaTilde})-(\ref{eq:CLNoXDotrdot}), the
time derivative of the candidate Lyapunov function $V$ can be written
as
\begin{multline*}
\dot{V}=-\tilde{\theta}^{T}Y^{T}\left(x\right)\tilde{x}-k\tilde{\theta}^{T}A_{s}\tilde{\theta}+k\tilde{\theta}^{T}Q_{s}-\alpha_{1}\tilde{x}^{T}\tilde{x}+r^{T}F\left(x,u\right)\tilde{\theta}-k\gamma_{1}r^{T}Y\left(x\right)\Gamma A_{s}\tilde{\theta}-r^{T}\gamma_{1}Y\left(x\right)\Gamma Y^{T}\left(x\right)\tilde{x}\\
+k\gamma_{1}r^{T}Y\left(x\right)\Gamma Q_{s}-k_{1}r^{T}r.
\end{multline*}
Provided the sufficient conditions in (\ref{eq:CLTNoXDotL1Suff})
are satisfied, the Lyapunov derivative can be bounded as
\[
\dot{V}\leq-\frac{k\underline{c}}{4}\left\Vert \tilde{\theta}\right\Vert ^{2}-\frac{\alpha_{1}}{3}\left\Vert \tilde{x}\right\Vert ^{2}-\frac{k_{1}}{8}\left\Vert r\right\Vert ^{2}+\left(\frac{k}{2\underline{a}}+\frac{k^{2}\overline{Y}^{2}\overline{\Gamma}^{2}}{k_{1}}\right)\left\Vert Q_{s}\right\Vert ^{2}.
\]
Using the hypothesis that the elements of the history stack are bounded
and the fact that $u$ is stabilizing, the Lyapunov derivative can
be bounded as
\[
\dot{V}\leq-\frac{v}{\overline{v}}V\left(Z\right)+\iota_{s}.
\]
Using the comparison lemma \cite[Lemma 3.4]{Khalil2002}, 
\[
V\left(Z\left(\tau\right)\right)\leq\left(\!\overline{V}_{s}-\frac{\overline{v}}{v}\iota_{s}\!\right)e^{-\frac{v}{\overline{v}}\left(\tau-t\right)}+\frac{\overline{v}}{v}\iota_{s},\:\forall\tau\!\in\!\left[t,t+T\right).
\]
If $\overline{V}_{s}\geq\frac{\overline{v}}{v}\iota_{s}$ then $V\left(Z\left(\tau\right)\right)\leq\overline{V}_{s}$.
If $\overline{V}_{s}<\frac{\overline{v}}{v}\iota_{s}$ then $V\left(Z\left(\tau\right)\right)\leq\frac{\overline{v}}{v}\iota_{s}.$
Hence, using the definition of $Z$ and the bounds in (\ref{eq:CLNoXDotVBounds}),
the bound in (\ref{eq:CLNoXDotThetaTildeBound}) is obtained.
\end{IEEEproof}

\subsubsection*{Part 2: Exponential decay of $\dot{\tilde{x}}$}

Let $Z_{r}\triangleq\left[\begin{array}{cc}
r^{T} & \tilde{x}^{T}\end{array}\right]^{T}\in\mathbb{R}^{2n}$ and let $V_{r}:\mathbb{R}^{2n}\to\mathbb{R}_{\geq0}$ be a candidate
Lyapunov function defined as $V_{r}\left(Z_{r}\right)\triangleq\left\Vert Z_{r}\right\Vert ^{2}.$
The following lemma establishes exponential convergence of the derivative
estimation error to a neighborhood of the origin using Lemma \ref{lem:CLTNoXDotThTBound}.
\begin{lem}
\label{lem:CLNoXDotXTildeBound}Let all the hypotheses of Lemma \ref{lem:CLTNoXDotThTBound}
be satisfied. Furthermore, assume temporarily that there exists a
constant $\overline{V}_{rs}$ such that the candidate Lyapunov function
$V_{r}$ satisfies $V_{r}\left(Z_{r}\left(t\right)\right)\leq\overline{V}_{rs}$.
Then, the Lyapunov function $V_{r}$ is bounded as 
\begin{equation}
V_{r}\!\left(\! Z_{r}\!\left(\tau\right)\!\right)\!\leq\!\left(\!\overline{V}_{rs}\!-\!\frac{\iota_{rs}}{v_{r}}\!\right)\! e^{-v_{r}\left(\tau-t\right)}\!+\!\frac{\iota_{rs}}{v_{r}},\:\forall\tau\!\in\!\left[t,t+T\right),\label{eq:CLNoXDotVrDecay}
\end{equation}
where $\nu_{r}=\min\left\{ \frac{k_{1}}{2},\alpha_{1}\right\} $ and
$\iota_{rs}\triangleq\frac{\left(\theta_{s}\overline{F}+k\overline{Y}\overline{\Gamma}\left(\theta_{s}\overline{A}+\overline{Q}_{s}\right)\right)^{2}}{k_{1}}.$
Furthermore, given a constant $\epsilon_{r}\in\mathbb{R}_{>0},$ the
gain $k_{1}$ can be selected large enough such that $\left\Vert \dot{\tilde{x}}\left(t+T\right)\right\Vert \leq\epsilon_{r}$.\end{lem}
\begin{IEEEproof}
Using (\ref{eq:CLNoXDotrdot}), the time derivative of the candidate
Lyapunov function $V_{r}$ can be written as 
\begin{multline*}
\dot{V}_{r}=-2r^{T}kY\left(x\right)\Gamma A_{s}\tilde{\theta}-2r^{T}Y\left(x\right)\Gamma Y^{T}\left(x\right)\tilde{x}+2r^{T}\left(\nabla Y\left(x\right)Y\left(x\right)\theta+\nabla Y\left(x\right)f_{1}\left(x\right)+\nabla Y\left(x\right)g\left(x\right)u\right)\tilde{\theta}\\
-2r^{T}\tilde{x}+k2r^{T}Y\left(x\right)\Gamma Q_{s}-k_{1}2r^{T}r+2\tilde{x}^{T}\left(r-\alpha_{1}\tilde{x}\right).
\end{multline*}
Completing the squares and using Lemma \ref{lem:CLTNoXDotThTBound},
the Lyapunov derivative can be bounded as 
\[
\dot{V}_{r}\leq-v_{r}V_{r}\left(Z_{r}\right)+\iota_{rs}.
\]
Using the comparison lemma, \cite[Lemma 3.4]{Khalil2002} 
\[
V_{r}\left(Z_{r}\left(\tau\right)\right)\leq\left(\!\overline{V}_{rs}-\frac{\iota_{rs}}{v_{r}}\!\right)\! e^{-v_{r}\left(\tau-t\right)}+\frac{\iota_{rs}}{v_{r}},\:\forall\tau\!\in\!\left[t,t+T\right).
\]
Using the fact that, $\dot{\tilde{x}}=r-\alpha_{1}\tilde{x}$ the
state-derivative estimation error can be bounded as 
\[
\left\Vert \dot{\tilde{x}}\right\Vert ^{2}\leq\left\Vert r\right\Vert ^{2}+\alpha_{1}\left\Vert \tilde{x}\right\Vert ^{2}\leq\left(1+\alpha_{1}\right)V_{r}\left(Z_{r}\right).
\]
Based on (\ref{eq:CLNoXDotVrDecay}), given $\overline{V}_{rs}\geq V_{r}\left(Z_{r}\left(t\right)\right)$,
$\epsilon_{r}>0$, the gain $k_{1}$ can be selected large enough
so that $V_{r}\left(Z_{r}\left(t+T\right)\right)\leq\frac{\epsilon_{r}^{2}}{\left(1+\alpha_{1}\right)}$.
Hence, given $\overline{V}_{rs},$ $\epsilon_{r}>0,$ the gain $k_{1}$
can be selected to be large enough so that $\left\Vert \dot{\tilde{x}}\left(t+T\right)\right\Vert \leq\epsilon_{r}.$
\end{IEEEproof}

\subsubsection*{Part 3: Asymptotic convergence to the origin}

Lemmas \ref{lem:CLTNoXDotThTBound} and \ref{lem:CLNoXDotXTildeBound}
employ the temporary hypothesis that the state-derivative estimates
$\dot{\hat{x}}_{j}$ stored in the history stack remain bounded. However,
since the estimates $\dot{\hat{x}}$ are generated dynamically using
(\ref{eq:CLNoXDotEst}), they can not be guaranteed to be bounded
a priori. In the following, the results of Lemmas \ref{lem:CLTNoXDotThTBound}
and \ref{lem:CLNoXDotXTildeBound} are used in an inductive argument
to show that all the states of the dynamical system defined by (\ref{eq:CLNoXDotThetaTilde})-(\ref{eq:CLNoXDotrdot})
remain bounded and decay to the origin asymptotically provided enough
data can be recorded to repopulate the history stack after each purge.%
\footnote{The case where the history stack can not be purged and repopulated
indefinitely is addressed in Section \ref{sub:CLNoXDotUltimate-boundedness-under}.%
}
\begin{thm}
\label{thm:CLNoXDotAsymptotic}Provided the history stacks $\mathcal{H}$
and $\mathcal{G}$ are populated using Algorithm \ref{alg:CLNoXDotpurgeDwell},
the learning gains are selected to satisfy $\overline{V}_{r1}>\frac{\iota_{r1}}{v_{r}}$
and the sufficient gain conditions in (\ref{eq:CLTNoXDotL1Suff}),
a bound $\overline{V}_{1}\in\mathbb{R}_{>0}$ is known such that $\overline{V}_{1}>\max\left(\frac{\overline{v}}{v}\iota_{1},V\left(Z\left(0\right)\right)\right)$,
and provided the system states are exciting such that the history
stack $\mbox{\ensuremath{\mathcal{H}}}$ can be persistently purged
and replenished, i.e., 
\begin{equation}
s\to\infty,\:\mbox{as}\quad t\to\infty,\label{eq:CLNoXDotPE}
\end{equation}
 then, $\left\Vert \tilde{\theta}\left(t\right)\right\Vert \to0$,
$\left\Vert r\left(t\right)\right\Vert \to0$, and $\left\Vert \tilde{x}\left(t\right)\right\Vert \to0$
as $t\to\infty$.\end{thm}
\begin{IEEEproof}
Let $\left\{ T_{s}\in\mathbb{R}_{\geq0}\mid s\in\mathbb{N}\right\} $
be a set of switching time instances defined as $T_{s}=\left\{ t\mid\rho\left(\tau\right)<s+1,\:\forall\tau\in\left[0,t\right)\land\rho\left(\tau\right)\geq s+1,\:\forall\tau\in\left[t,\infty\right)\right\} .$
That is, for a given switching index $s,$ $T_{s}$ denotes the time
instance when the $\left(s+1\right)$\textsuperscript{th} subsystem
is switched on. To facilitate proof by mathematical induction, assume
temporarily that the hypotheses of Lemmas \ref{lem:CLTNoXDotThTBound}
and \ref{lem:CLNoXDotXTildeBound} are satisfied for $t\in\left[0,T_{s}\right)$
for some $s$. Furthermore, assume temporarily that the following
sufficient condition is satisfied: 
\begin{equation}
\iota_{rj}>\iota_{r\left(j+1\right)},\quad\iota_{j}>\iota_{j+1},\quad\forall j\in\left\{ 1,2,\cdots,s-1\right\} ,\label{eq:CLNoXDotTh1SuffCond}
\end{equation}
Then, using (\ref{eq:CLNoXDotVDecay}) and (\ref{eq:CLNoXDotVrDecay}),
the Lyapunov functions $V$ and $V_{r}$ can be bounded as $V\left(Z\left(t\right)\right)\leq W\left(t\right)\triangleq\left(\overline{V}_{1}-\frac{\overline{v}}{v}\iota_{1}\right)e^{-\frac{v}{\overline{v}}\left(t\right)}+\frac{\overline{v}}{v}\iota_{s}+\sum_{j=1}^{s-1}\frac{\overline{v}}{v}\left(\iota_{j}-\iota_{j+1}\right)e^{-\frac{v}{\overline{v}}\left(t-T_{j}\right)},$
and $V_{r}\left(Z_{r}\left(t\right)\right)\leq W_{r}\left(t\right)\triangleq\left(\overline{V}_{r1}-\frac{\iota_{r1}}{v_{r}}\right)e^{-v_{r}\left(t\right)}+\frac{\iota_{rs}}{v_{r}}+\sum_{j=1}^{s-1}\left(\frac{\iota_{rj}}{v_{r}}-\frac{\iota_{r\left(j+1\right)}}{v_{r}}\right)e^{-v_{r}\left(t-T_{j}\right)},$
where the constants $\iota_{s}$ and $\iota_{rs}$ were introduced
in (\ref{eq:CLNoXDotVDecay}) and (\ref{eq:CLNoXDotVrDecay}), respectively,
and $W$, and $W_{r}$ denote the envelopes that bound the Lyapunov
functions $V$ and $V_{r}$, respectively. Using the bounds on the
Lyapunov functions, the bounding envelopes at consecutive switching
instances can be related as
\begin{multline*}
W\left(T_{s}\right)-W\left(T_{s-1}\right)=\frac{\overline{v}}{v}\left(\iota_{s-1}-\iota_{s}\right)\left(e^{\frac{v}{\overline{v}}\left(T_{s-1}-T_{s}\right)}-1\right)+\sum_{j=1}^{s-2}\frac{\overline{v}}{v}\left(\iota_{j}-\iota_{j+1}\right)e^{-\frac{v}{\overline{v}}\left(T_{s}-T_{j}\right)}\left(1-e^{\frac{v}{\overline{v}}\left(T_{s}-T_{s-1}\right)}\right)\\
+\left(\overline{V}_{1}-\frac{\overline{v}}{v}\iota_{1}\right)\left(e^{-\frac{v}{\overline{v}}\left(T_{s}\right)}-e^{-\frac{v}{\overline{v}}\left(T_{s-1}\right)}\right),
\end{multline*}
and 
\begin{multline*}
W_{r}\left(T_{s}\right)\!-\! W_{r}\left(T_{s-1}\right)\!=\!\left(\!\frac{\iota_{r\left(s-1\right)}}{v_{r}}-\frac{\iota_{rs}}{v_{r}}\!\right)\!\left(\! e^{v_{r}\left(T_{s-1}-T_{s}\right)}\!-\!1\!\right)+\sum_{j=1}^{s-2}\left(\frac{\iota_{rj}}{v_{r}}-\frac{\iota_{r\left(j+1\right)}}{v_{r}}\right)e^{-v_{r}\left(T_{s}-T_{j}\right)}\left(1-e^{v_{r}\left(T_{s}-T_{s-1}\right)}\right)\\
+\left(\overline{V}_{r1}-\frac{\iota_{r1}}{v_{r}}\right)\left(e^{-v_{r}\left(T_{s}\right)}-e^{-v_{r}\left(T_{s-1}\right)}\right).
\end{multline*}
Since $T_{s}>T_{s-1},$ the terms $\left(e^{-\frac{v}{\overline{v}}\left(T_{s}\right)}-e^{-\frac{v}{\overline{v}}\left(T_{s-1}\right)}\right)$,
$\left(1-e^{v_{r}\left(T_{s}-T_{s-1}\right)}\right)$, $\left(1-e^{v_{r}\left(T_{s}-T_{s-1}\right)}\right),$
$\left(e^{\frac{v}{\overline{v}}\left(T_{s-1}-T_{s}\right)}-1\right)$,
$\left(e^{-v_{r}\left(T_{s}\right)}-e^{-v_{r}\left(T_{s-1}\right)}\right)$,
and $\left(e^{v_{r}\left(T_{s-1}-T_{s}\right)}-1\right)$ are always
negative. By selecting $\overline{V}_{1}$ as 
\begin{equation}
\overline{V}_{1}>\max\Biggl(\frac{\overline{v}}{v}\iota_{1},\overline{v}\biggl(\overline{F}_{1}^{2}+\overline{\theta}^{2}+\left\Vert \dot{\hat{x}}\left(0\right)\right\Vert ^{2}+\left(1+\alpha_{1}\right)^{2}\left\Vert \tilde{x}\left(0\right)\right\Vert ^{2}+\left\Vert \hat{\theta}\left(0\right)\right\Vert ^{2}\biggr)\Biggr)\label{eq:CLNoXDotV_1bar}
\end{equation}
and using (\ref{eq:CLNoXDotTh1SuffCond}) and the hypotheses of Theorem
\ref{thm:CLNoXDotAsymptotic}, then $W\left(T_{s}\right)<W\left(T_{s-1}\right)$
and $W_{r}\left(T_{s}\right)<W_{r}\left(T_{s-1}\right)$.

Since the history stack $\mathcal{H}_{1}$ is selected at random to
include bounded elements, all the hypotheses of Lemmas \ref{lem:CLTNoXDotThTBound}
and \ref{lem:CLNoXDotXTildeBound} are satisfied over the time interval
$\left[0,T_{1}\right)$. Hence, $V\left(Z\left(t\right)\right)\leq W\left(t\right)=\left(\overline{V}_{1}-\frac{\overline{v}}{v}\iota_{1}\right)e^{-\frac{v}{\overline{v}}\left(t\right)}+\frac{\overline{v}}{v}\iota_{1}$,
where $\iota_{1}=\left(\frac{k}{2\underline{a}}+\frac{k^{2}\overline{Y}^{2}\overline{\Gamma}^{2}}{k_{1}}\right)\overline{Q}_{1}^{2}$.
Using the bounds in (\ref{eq:CLNoXDotBounds}), $\overline{Q}_{1}$
can be computed as $\overline{Q}_{1}=qM\overline{Y}\left(\overline{F}_{1}+\overline{H}_{1}\right)$,
where $q>1$ is an adjustable parameter. Furthermore, $V_{r}\left(Z_{r}\left(t\right)\right)\leq W_{r}\left(t\right)=\left(\overline{V}_{r1}-\frac{\iota_{r1}}{v_{r}}\right)e^{-v_{r}\left(t\right)}+\frac{\iota_{r1}}{v_{r}}$,
where $\iota_{r1}=\frac{\left(\theta_{1}\overline{F}+k\overline{Y}\overline{\Gamma}\left(\theta_{1}\overline{A}+\overline{Q}_{1}\right)\right)^{2}}{k_{1}}$
and $\theta_{1}=\sqrt{\frac{\overline{v}}{\underline{v}}}\max\left\{ \sqrt{\overline{V}_{1}},\sqrt{\iota_{1}}\right\} .$
Using the sufficient conditions $\overline{V}_{1}>\frac{\overline{v}}{v}\iota_{1}$
and $\overline{V}_{r1}>\frac{\iota_{r1}}{v_{r}}$ stated in Theorem
\ref{thm:CLNoXDotAsymptotic}, it can be concluded that $W\left(T_{1}\right)<W\left(0\right)$
and $W_{r}\left(T_{1}\right)<W_{r}\left(0\right)$. Selecting $\overline{V}_{2}=W\left(T_{1}\right)$
and $\overline{V}_{r2}=W_{r}\left(T_{1}\right)$, it can be concluded
that $\overline{V}_{2}<\overline{V}_{1}$ and $\overline{V}_{r2}<\overline{V}_{r1}$.

Since the Lyapunov function $V_{r}$ does not grow beyond its initial
condition over the time interval $\left[0,T_{1}\right)$, $\sup_{t\in\left[0,T_{1}\right)}\left\Vert \dot{\tilde{x}}\left(t\right)\right\Vert \leq\left(1+\alpha_{1}\right)\overline{V}_{r1}$.
Moreover, since the history stack $\mathcal{H}_{2}$, which is active
over the time interval $\left[T_{1},T_{2}\right)$, is recorded over
the time interval $\left[0,T_{1}\right),$ all the hypotheses of Lemmas
\ref{lem:CLTNoXDotThTBound} and \ref{lem:CLNoXDotXTildeBound} are
also satisfied over the time interval $\left[T_{1},T_{2}\right)$,
and the constant $\overline{Q}_{2}$ can be computed as $\overline{Q}_{2}=M\overline{Y}\sqrt{2\left(1+\alpha_{1}\right)\overline{V}_{r1}}$.
Provided $q$ is selected such that $q>\max\left\{ 1,\frac{\sqrt{2\left(1+\alpha_{1}\right)\overline{V}_{r1}}}{\left(\overline{F}_{1}+\overline{H}_{1}\right)}\right\} ,$
then $\overline{Q}_{2}<\overline{Q}_{1}.$ Hence, $\iota_{2}<\iota_{1}$.
Since $\iota_{2}<\iota_{1}$ and $\overline{V}_{2}<\overline{V}_{1}$,
then $\theta_{2}<\theta_{1},$ and hence, $\iota_{r2}<\iota_{r1}.$ 

Hence, by mathematical induction, the hypotheses of Lemmas \ref{lem:CLTNoXDotThTBound}
and \ref{lem:CLNoXDotXTildeBound} are satisfied for all $t\in\mathbb{R}_{\geq0}$,
and $V\left(Z\left(T_{s}\right)\right)<V\left(Z\left(T_{s-1}\right)\right)$
for all $s\in\mathbb{N}$. Hence, $V\left(Z\left(t\right)\right)\to0$
as $t\to\infty.$ Since the Lyapunov function $V$ is common among
all the subsystems, $Z\left(t\right)\to0$ as $t\to\infty$
\end{IEEEproof}
Theorem \ref{thm:CLNoXDotAsymptotic} implies that provided the system
states are persistently excited such that the history stack $\mathcal{H}$
can always be replaced with a new full rank history stack, the state-derivative
estimates, and the parameter estimate vector asymptotically converge
to the state-derivative and the ideal parameter vector, respectively.
However, from a practical perspective, it may be undesirable for a
system to be in a persistently exciting state, or excitation beyond
a certain finite time-interval may not be available. If excitation
is available only over a finite time-interval, then the parameter
estimation errors can be shown to be uniformly ultimately bounded,
provided the history stacks are updated so that the time interval
between two consecutive updates, i.e., the dwell time, is large enough.
Algorithm \ref{alg:CLNoXDotpurgeDwell} guarantees a minimum dwell
time between two consecutive updates provided the constant $\mathcal{T}$
is selected large enough.

\subsection{\label{sub:CLNoXDotUltimate-boundedness-under}Ultimate boundedness
under finite excitation}

For notational brevity, let $\beta_{1}\triangleq\left(\theta_{1}\overline{F}+k\overline{Y}\overline{\Gamma}\left(\theta_{1}\overline{A}+\overline{Q}_{1}\right)\right)^{2}$
and $\beta_{2}\triangleq\left(\frac{k}{2\underline{a}}+\frac{k^{2}\overline{Y}^{2}\overline{\Gamma}^{2}}{k_{1}}\right)$.
\begin{thm}
\label{thm:CLNoXDotUUB}Let all the hypotheses of Theorem \ref{thm:CLNoXDotAsymptotic}
be satisfied, except for the persistent excitation hypothesis in (\ref{eq:CLNoXDotPE}).
Let $T\in\mathbb{R}_{>0}$ be a time instance such that $T_{s}<T$
for some $s>2$. Provided the history stacks are updated and populated
using Algorithm \ref{alg:CLNoXDotpurgeDwell}, where the minimum dwell
time satisfies
\[
\mathcal{T}\geq\max_{j\in\left\{ 1,2,\cdots,s-1\right\} }\Biggl\{\frac{\overline{v}}{v\left(s-j\right)}\log\left(\frac{s\iota_{j}}{\iota_{s}}-\frac{s\iota_{j+1}}{\iota_{s}}\right),\frac{\overline{v}}{vs}\log\left(\frac{sv\overline{V}_{1}}{\overline{v}\iota_{s}}-\frac{s\iota_{1}}{\iota_{s}}\right),\frac{1}{v_{r}}\log\left(\frac{v_{r}\overline{V}_{r1}}{\iota_{r1}}-1\right)\Biggr\},
\]
and the sufficient gain condition $k_{1}>\frac{8v_{r}\overline{v}^{2}M\overline{Y}\left(1+\alpha_{1}\right)\beta_{1}\beta_{2}^{2}}{\underline{v}^{2}v^{2}\epsilon^{2}}$
is satisfied for a given $\epsilon>0,$ then $\sup_{t>T}\left\Vert \tilde{\theta}\left(t\right)\right\Vert \leq\epsilon$.\end{thm}
\begin{IEEEproof}
Since the history stack $\mathcal{H}_{1}$ is selected at random,
all the hypotheses of Lemmas \ref{lem:CLTNoXDotThTBound} and \ref{lem:CLNoXDotXTildeBound}
are satisfied over the time interval $\left[0,T_{1}\right)$. Hence,
$V\left(Z\left(t\right)\right)\leq W\left(t\right)=\left(\overline{V}_{1}-\frac{\overline{v}}{v}\iota_{1}\right)e^{-\frac{v}{\overline{v}}\left(t\right)}+\frac{\overline{v}}{v}\iota_{1}$,
where $\iota_{1}=\beta_{2}\overline{Q}_{1}^{2}$. Using the bounds
in (\ref{eq:CLNoXDotBounds}), $\overline{Q}_{1}$ can be computed
as $\overline{Q}_{1}=qM\overline{Y}\left(\overline{F}_{1}+\overline{H}_{1}\right)$,
where $q>1$ is an adjustable parameter. Furthermore, $V_{r}\left(Z\left(t\right)\right)\leq W_{r}\left(t\right)=\left(\overline{V}_{r1}-\frac{\iota_{r1}}{v_{r}}\right)e^{-v_{r}\left(t\right)}+\frac{\iota_{r1}}{v_{r}}$,
where $\iota_{r1}=\frac{\beta_{1}}{k_{1}}$ and $\theta_{1}=\sqrt{\frac{\overline{v}}{\underline{v}}}\max\left\{ \sqrt{\overline{V}_{1}},\sqrt{\iota_{1}}\right\} .$
Provided $T_{1}>\frac{1}{v_{r}}\log\left(\frac{v_{r}\overline{V}_{r1}}{\iota_{r1}}-1\right)$,
then $W_{r}\left(T_{1}\right)\leq\frac{2\iota_{r1}}{v_{r}}.$ 

Let $\epsilon_{1}>0$ be a constant, to be selected later. Provided
the gain $k_{1}$ is selected such that $k_{1}>\frac{2v_{r}\left(1+\alpha_{1}\right)\beta_{1}}{\epsilon_{1}^{2}}$,
then $\overline{V}_{r2}\leq\frac{\epsilon_{1}^{2}}{\left(1+\alpha_{1}\right)}$.
Since $T_{2}<T$ by hypothesis, the Lyapunov envelope $W_{r}$ decays
over the time interval $\left[T_{1},T_{2}\right)$; hence, $\sup_{t\in\left[T_{1},T_{2}\right)}\left\Vert \dot{\tilde{x}}\left(t\right)\right\Vert \leq\epsilon_{1}$,
which implies $\overline{Q}_{3}$ can be selected as $\overline{Q}_{3}=M\overline{Y}\epsilon_{1}$
and $\iota_{3}=\beta_{2}M\overline{Y}\epsilon_{1}.$ Selecting $\epsilon_{1}=\frac{\underline{v}v\epsilon}{2\overline{v}\beta_{2}M\overline{Y}},$
the inequality $\iota_{3}\leq\frac{v\underline{v}\epsilon}{2\overline{v}}$
is obtained.

If the history stack $\mathcal{H}$ is not updated after the time
instance $T_{s}$, then using an inductive argument similar to the
proof of Theorem \ref{thm:CLNoXDotAsymptotic}, it can be concluded
that all the hypotheses of Lemma \ref{lem:CLTNoXDotThTBound} are
satisfied for all $t\in\left[T_{s},\infty\right)$. Hence, the Lyapunov
function $V$ is bounded for all $t\in\left[T_{s},\infty\right)$
by 
\[
V\left(Z\left(t\right)\right)\leq\left(\overline{V}_{1}-\frac{\overline{v}}{v}\iota_{1}\right)e^{-\frac{v}{\overline{v}}\left(T_{s}\right)}+\frac{\overline{v}}{v}\iota_{s}+\sum_{j=1}^{s-1}\frac{\overline{v}}{v}\left(\iota_{j}-\iota_{j+1}\right)e^{-\frac{v}{\overline{v}}\left(T_{s}-T_{j}\right)}.
\]
Since Algorithm \ref{alg:CLNoXDotpurgeDwell} is designed to allow
a minimum dwell time of $\mathcal{T}$ seconds, $T_{s}\geq s\mathcal{T}$
and $T_{s}-T_{j}\geq\left(s-j\right)\mathcal{T}$. Hence, 
\[
V\left(Z\left(T_{s}\right)\right)\leq\left(\overline{V}_{1}-\frac{\overline{v}}{v}\iota_{1}\right)e^{-\frac{v}{\overline{v}}\left(s\mathcal{T}\right)}+\frac{\overline{v}}{v}\iota_{s}+\sum_{j=1}^{s-1}\frac{\overline{v}}{v}\left(\iota_{j}-\iota_{j+1}\right)e^{-\frac{v}{\overline{v}}\mathcal{T}\left(s-j\right)}.
\]
Provided the dwell time satisfies $\frac{\overline{v}}{vs}\log\left(\frac{sv\overline{V}_{1}}{\overline{v}\iota_{s}}-\frac{s\iota_{1}}{\iota_{s}}\right)\leq\mathcal{T},$
and $\frac{\overline{v}}{v\left(s-j\right)}\log\left(\frac{s\iota_{j}}{\iota_{s}}-\frac{s\iota_{j+1}}{\iota_{s}}\right)\leq\mathcal{T},\:\forall j\in\left\{ 1,2,\cdots,s-1\right\} ,$
then $V\left(Z\left(t\right)\right)\leq\frac{2\overline{v}\iota_{s}}{v}\leq\frac{2\overline{v}\iota_{3}}{v}$,
for all $t\in\left[T_{s},\infty\right)$. Hence, using the bounds
on the Lyapunov function in (\ref{eq:CLNoXDotVBounds}), $\left\Vert \tilde{\theta}\left(t\right)\right\Vert \leq\frac{2\overline{v}\iota_{3}}{\underline{v}v}$
for all $t\in\left[T_{s},\infty\right)$. Using the bound $\iota_{3}\leq\frac{v\underline{v}\epsilon}{2\overline{v}}$,
$\left\Vert \tilde{\theta}\left(t\right)\right\Vert \leq\epsilon$
for all $t\in\left[T_{s},\infty\right)$.
\end{IEEEproof}

\section{Simulation results}

The developed technique is simulated using a model for a two-link
robot manipulator arm. The four-dimensional state of the model is
denoted by $x\triangleq\begin{bmatrix}x_{1} & x_{2} & x_{3} & x_{4}\end{bmatrix}^{T}$.
The dynamics of the model are described by (\ref{eq:CLNoXDotDyn}),
where
\begin{equation}
f_{1}\left(x\right)=\begin{bmatrix}x_{3}\\
x_{4}\\
-\left(M\left(x\right)\right)^{-1}V_{m}\begin{bmatrix}x_{3}\\
x_{4}
\end{bmatrix}
\end{bmatrix},G\left(x\right)=\begin{bmatrix}\begin{array}{cc}
0 & 0\end{array}\\
\begin{array}{cc}
0 & 0\end{array}\\
\left(M\left(x\right)\right)^{-1}
\end{bmatrix},Y\left(x\right)=\begin{bmatrix}\begin{array}{cccc}
0 & 0 & 0 & 0\end{array}\\
\begin{array}{cccc}
0 & 0 & 0 & 0\end{array}\\
\left[\begin{array}{cc}
\left(M\left(x\right)\right)^{-1} & \left(M\left(x\right)\right)^{-1}\end{array}\right]D\left(x\right)
\end{bmatrix}.\label{eq:CLNoXDotSymDyn}
\end{equation}
In (\ref{eq:CLNoXDotSymDyn}), $D\left(x\right)\triangleq\mbox{diag}\left[x_{3},\: x_{4},\:\tanh\left(x_{3}\right),\:\tanh\left(x_{4}\right)\right]$,
$M\left(x\right)\triangleq\begin{bmatrix}p_{1}+2p_{3}c_{2}\left(x\right), & p_{2}+p_{3}c_{2}\left(x\right)\\
p_{2}+p_{3}c_{2}\left(x\right), & p_{2}
\end{bmatrix},$ and $V_{m}\left(x\right)\triangleq\begin{bmatrix}-p_{3}s_{2}\left(x\right)x_{4}, & -p_{3}s_{2}\left(x\right)\left(x_{3}+x_{4}\right)\\
p_{3}s_{2}\left(x\right)x_{3}, & 0
\end{bmatrix},$ where $c_{2}\left(x\right)=cos\left(x_{2}\right),$ $s_{2}\left(x\right)=sin\left(x_{2}\right)$,
and $p_{1}=3.473$, $p_{2}=0.196$, and $p_{3}=0.242$ are constants.
The system has four unknown parameters. The ideal values of the unknown
parameters are $\theta=\begin{bmatrix}5.3 & 1.1 & 8.45 & 2.35\end{bmatrix}^{T}.$ 

The developed technique is compared against numerical differentiation-based
concurrent learning where the numerical derivatives are computed using
polynomial regression over a window of collected data. The state measurements
are filtered using a moving average filter for  state-derivative estimation.
To facilitate the comparison, multiple simulation runs are performed
using a combination of gains, window sizes, and thresholds for two
levels of noise. The low-noise and high-noise simulations are performed
by adding white Gaussian noise with variance 0.005 and 0.1, respectively,
to the state measurements. The simulations are repeated five times
for each combination of gains, and the gains, window sizes, and thresholds
that yield the lowest steady-state RMS error over five runs are selected
for comparison. Table \ref{tab:CLNoXDotCompare} indicates that the
developed technique outperforms numerical differentiation-based concurrent
learning for both low-noise and high-noise cases. The following figures
illustrate the performance of the developed technique in one sample
run.
\begin{table}
\begin{centering}
\begin{tabular}{>{\centering}m{0.17\columnwidth}>{\centering}m{0.15\columnwidth}>{\centering}m{0.15\columnwidth}>{\centering}m{0.15\columnwidth}>{\centering}m{0.15\columnwidth}}
\hline 
 & \multicolumn{2}{>{\centering}m{0.3\columnwidth}}{Numerical differentiation-based CL} & \multicolumn{2}{>{\centering}m{0.3\columnwidth}}{Developed technique}\tabularnewline
\hline 
\hline 
Noise variance & $0.005$ & $0.1$ & $0.005$ & $0.1$\tabularnewline
\hline 
\hline 
RMS steady-state error & $1.75$ & $17.55$ & $0.27$ & $0.46$\tabularnewline
\hline 
\end{tabular}
\par\end{centering}

\protect\caption{\label{tab:CLNoXDotCompare}Simulation results for the developed technique
and numerical differentiation-based CL.}
\end{table}

\begin{figure}
\subfloat[\label{fig:CLNoXDot2Lx}State trajectory.]{\noindent \begin{centering}
\includegraphics[width=0.5\columnwidth]{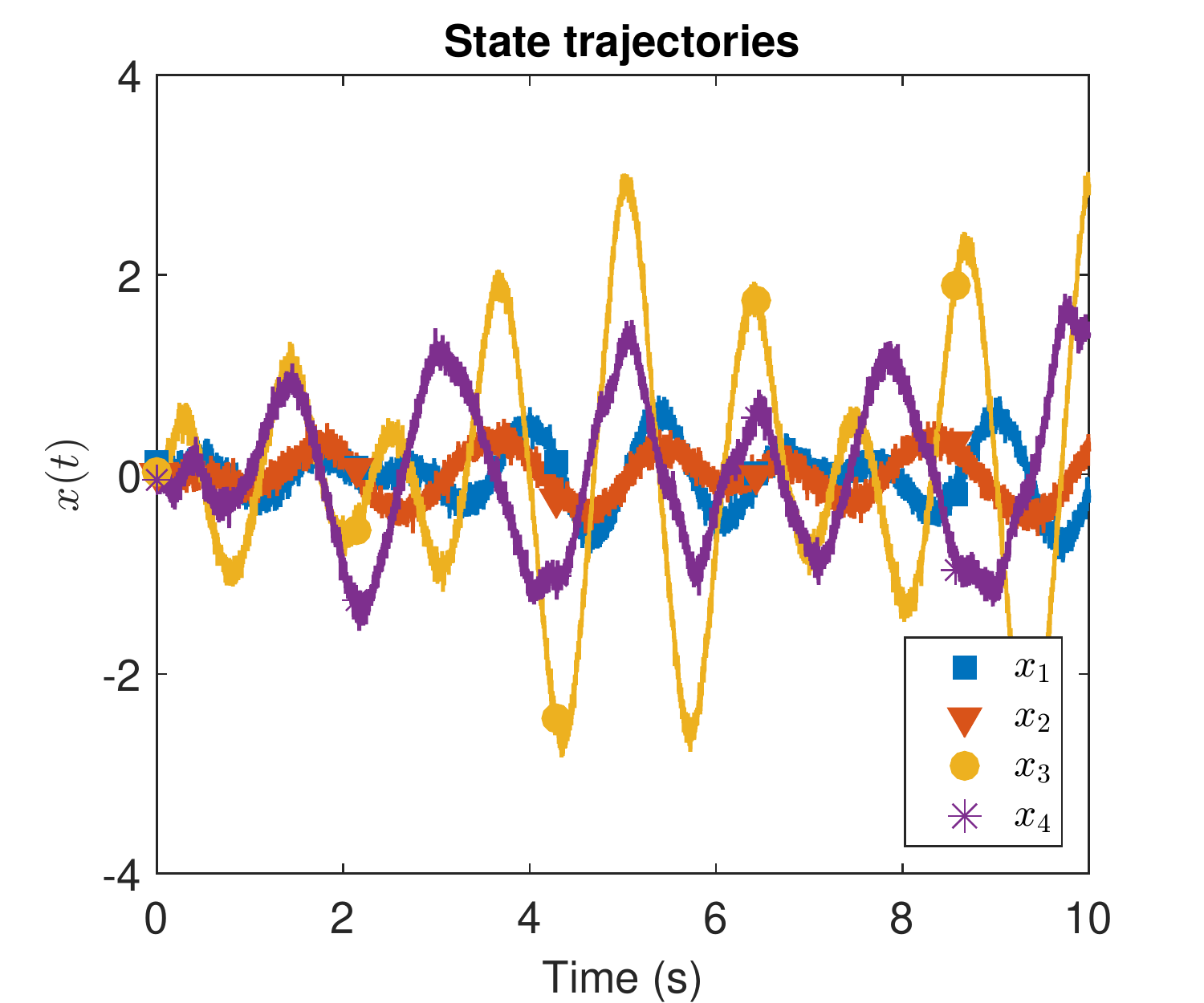}
\par\end{centering}

}\subfloat[\label{fig:CLNoXDot2LthetaH}Trajectories of the parameter estimates.
Dashed lines represent the ideal values. ]{\noindent \begin{centering}
\includegraphics[width=0.5\columnwidth]{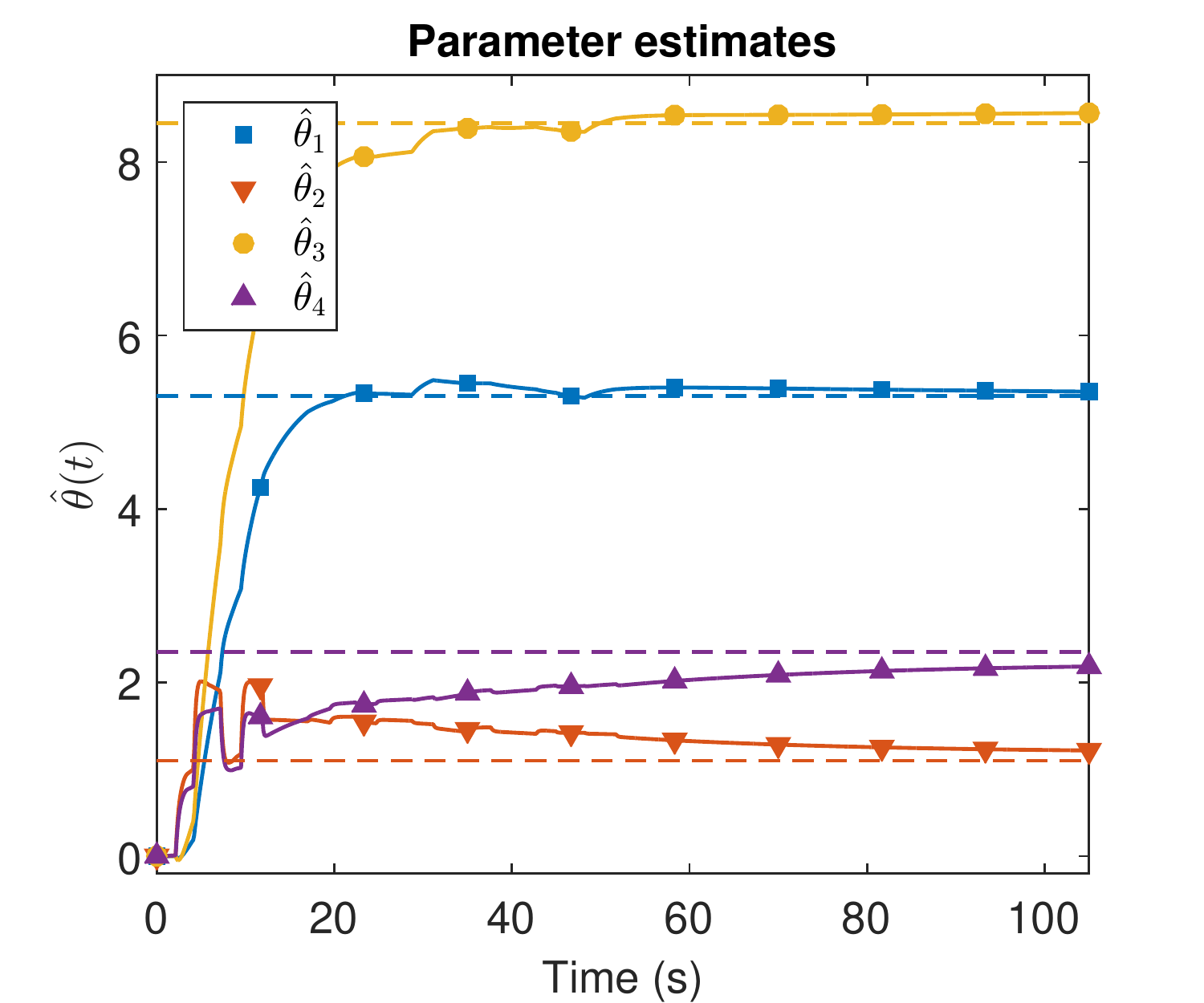}
\par\end{centering}

}

\protect\caption{Trajectories of the system state and the parameter estimates.}
\end{figure}

\begin{figure}
\subfloat[\label{fig:CLNoXDot2LxT}State estimation error.]{\noindent \begin{centering}
\includegraphics[width=0.5\columnwidth]{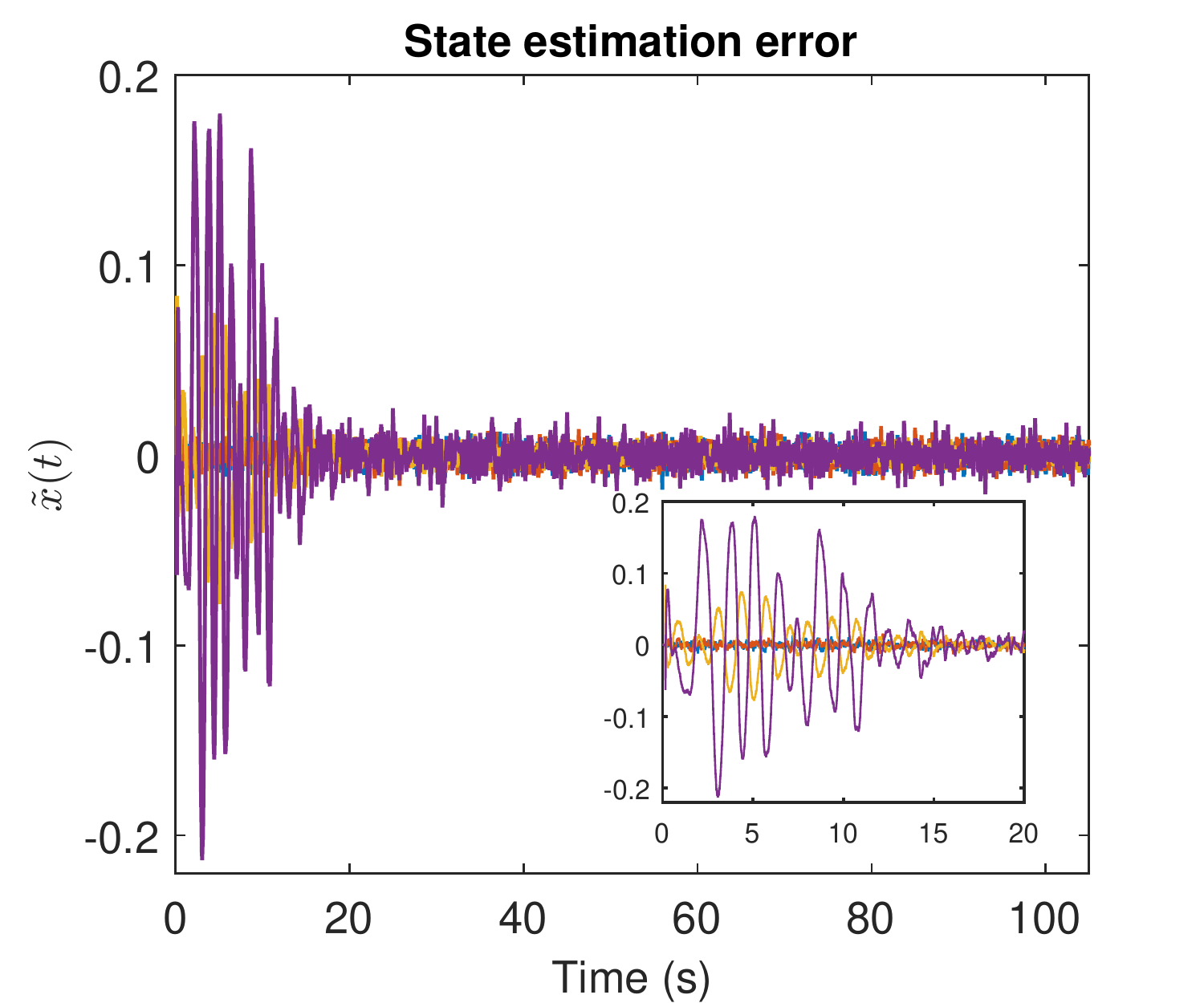}
\par\end{centering}

}\subfloat[\label{fig:CLNoXDot2LxTD}State-derivative estimation error.]{\noindent \begin{centering}
\includegraphics[width=0.5\columnwidth]{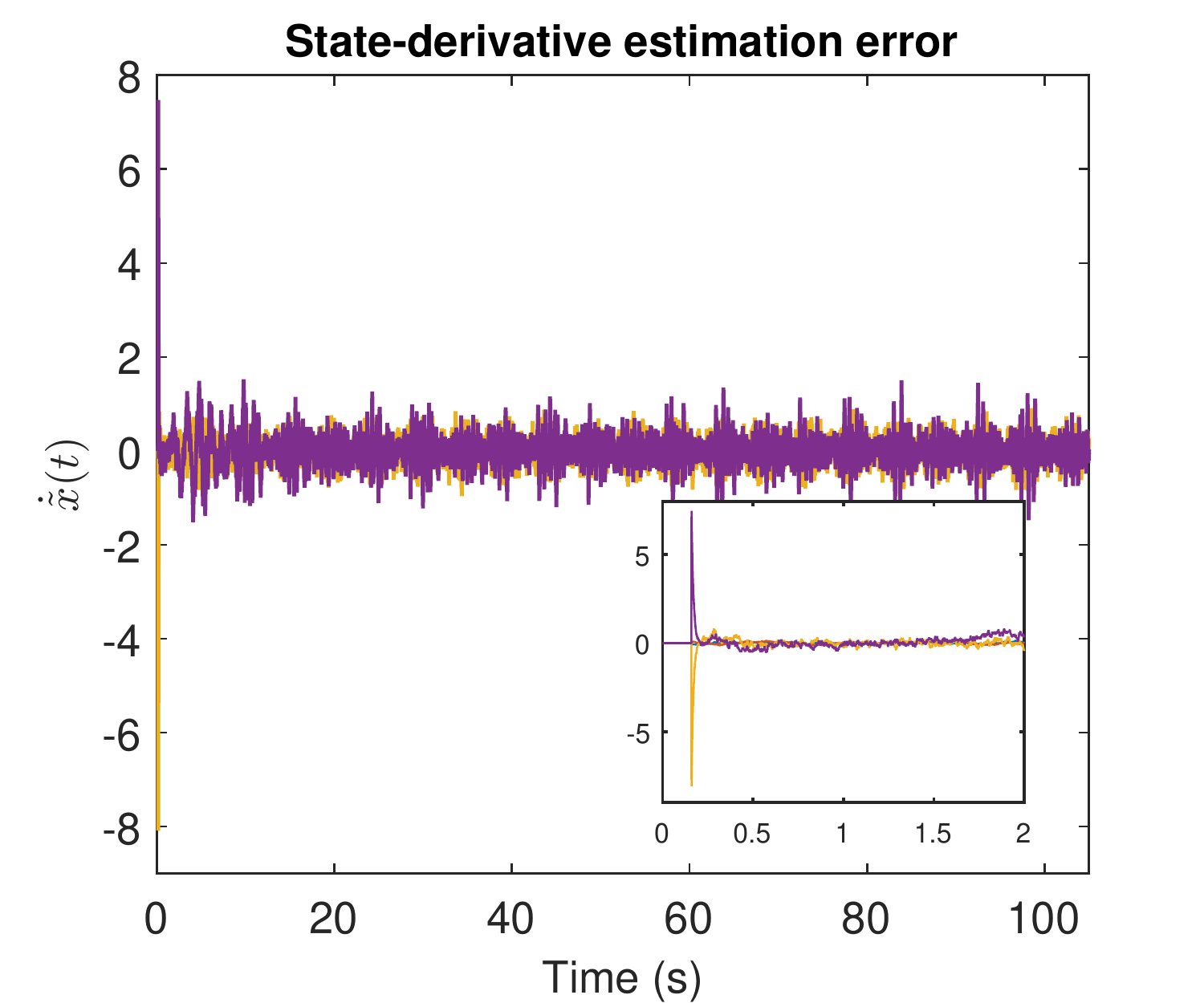}
\par\end{centering}

}

\protect\caption{Performance of the  state-derivative estimator.}
\end{figure}

\begin{figure}
\subfloat[\label{fig:CLNoXDot2LSingValue}Minimum singular value of the history
stack.]{\noindent \begin{centering}
\includegraphics[width=0.5\columnwidth]{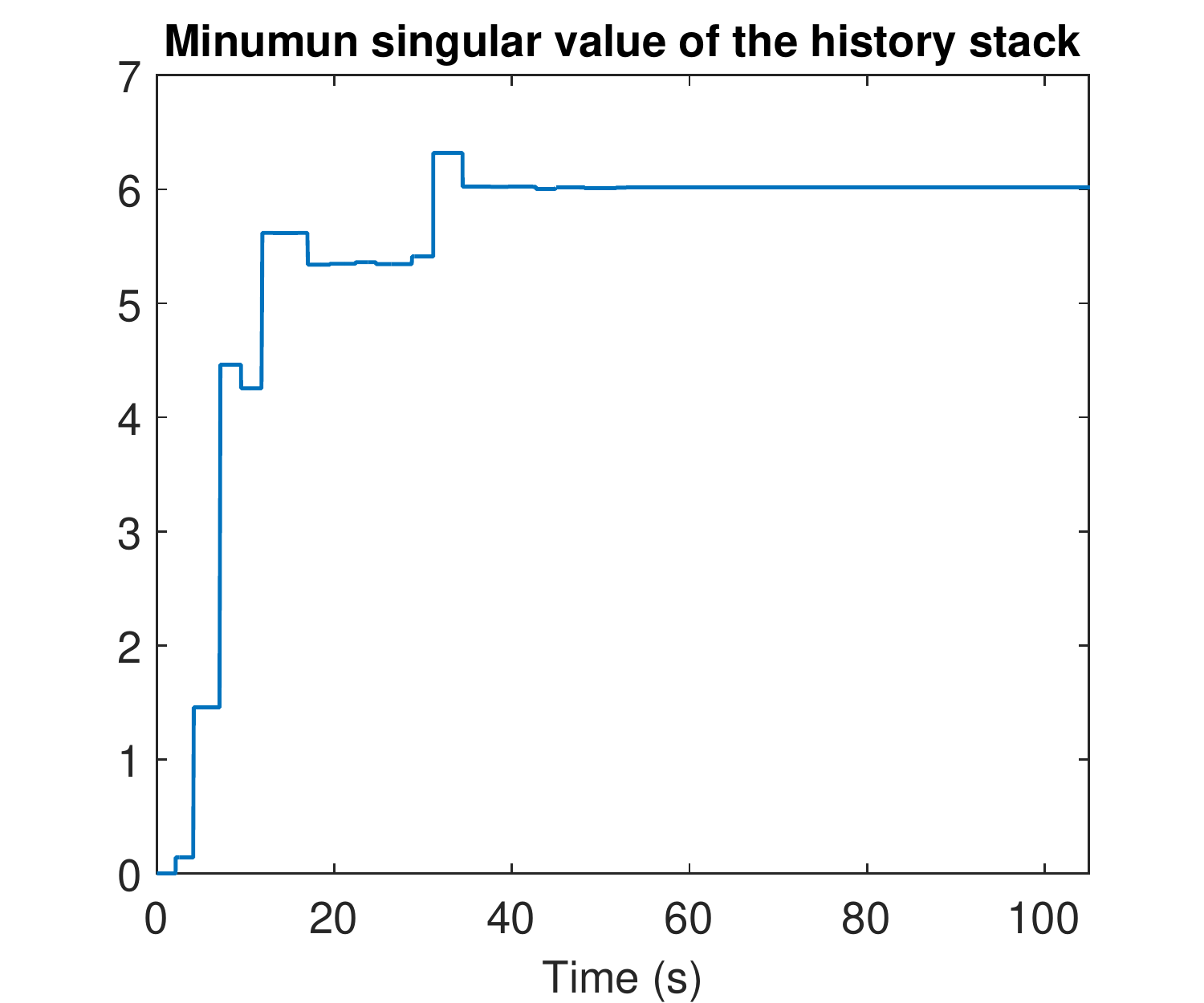}
\par\end{centering}

}\subfloat[\label{fig:CLNoXDot2LPurgeIndex}Purging index]{\noindent \begin{centering}
\includegraphics[width=0.5\columnwidth]{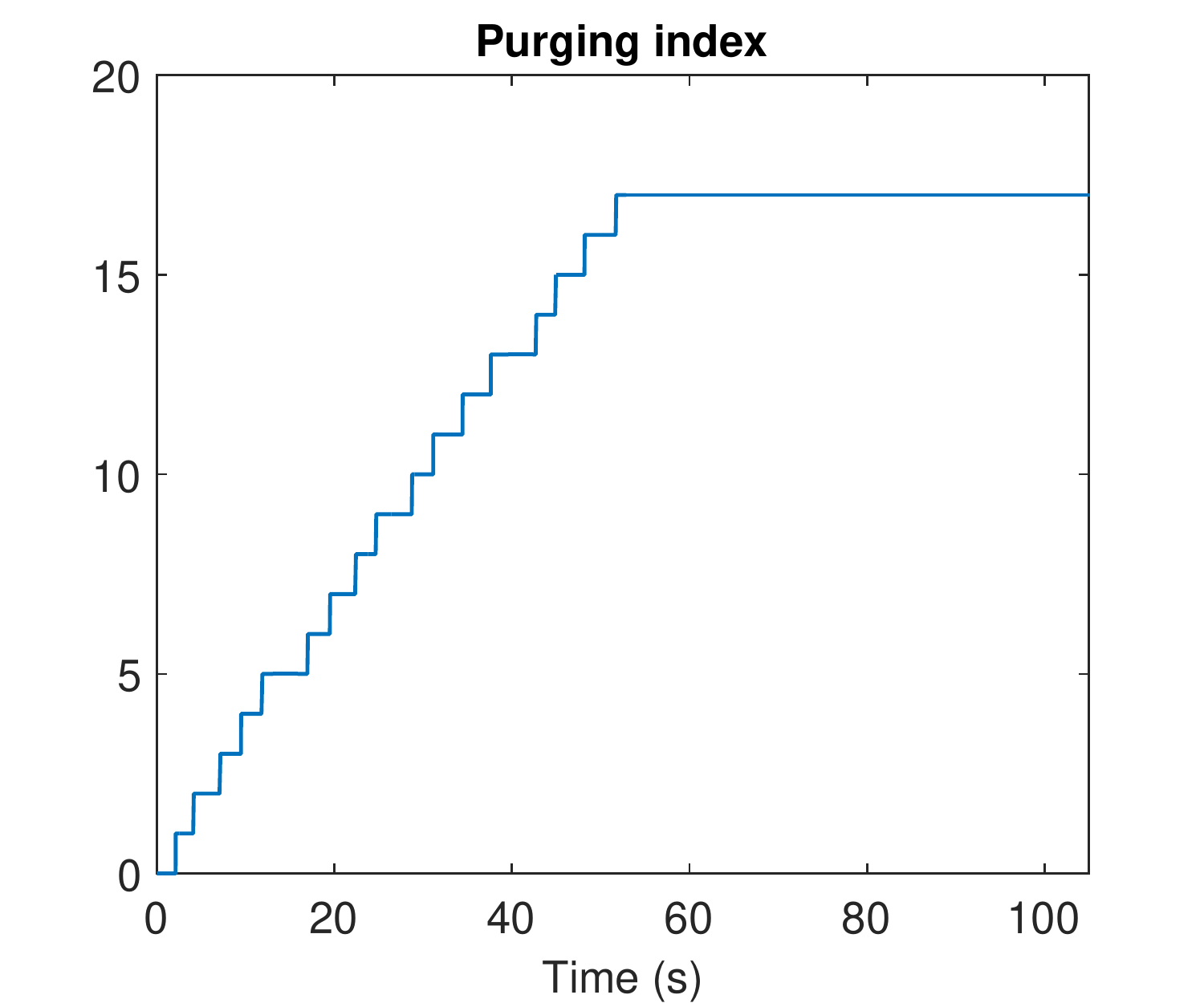}
\par\end{centering}

}

\protect\caption{Purging index and the minimum singular value for the history stack.}
\end{figure}

Figure \ref{fig:CLNoXDot2Lx} shows the evolution of the system state,
where the added noise signal can be observed. Figure \ref{fig:CLNoXDot2LthetaH}
demonstrates convergence of the unknown parameters to a neighborhood
of their true values, where the dashed lines represent the true values.
Figure \ref{fig:CLNoXDot2LxT} shows the convergence of the state
estimation error to a ball around the origin. Figure \ref{fig:CLNoXDot2LxTD}
shows the convergence of the state-derivative estimation error to
a ball around the origin. The transients in Figures \ref{fig:CLNoXDot2LxT}
and \ref{fig:CLNoXDot2LxTD} necessitate the need for history stack
purging. Figure \ref{fig:CLNoXDot2LSingValue} shows the minimum singular
value of the history stack $\mathcal{H}$. The singular value is increasing
because of the thresholding algorithm. In this simulation, the threshold
parameter $\xi$ is set to one. Figure \ref{fig:CLNoXDot2LPurgeIndex}
shows the increments of the purging index, It can be observed that
the history stack gets purged faster initially as transients offer
significant data, and then the rate of purging levels off approximately
to a constant as the system achieves steady state. %
\footnote{The measurement noise does not inject excitation into the system since
the noisy measurements are used only for parameter estimation, and
the true state is used for feedback control.%
}

\section{Concluding remarks}

A concurrent learning-based parameter estimator is developed for a
linearly parameterized control-affine nonlinear system. An adaptive
observer is employed to generate the state-derivative estimates required
for concurrent learning. The developed technique is validated via
simulations on a nonlinear system where the state measurements are
corrupted by additive Gaussian noise. The simulation results indicate
that the developed technique yields better results than numerical
differentiation-based CL, even more so as the variance of the additive
noise is increased. Even though the simulation results indicate a
degree of robustness to measurement noise, the theoretical development
does not account for measurement noise. 

Measurement noise affects the developed parameter estimator in two
ways. An error is introduced in the state-derivative estimates generated
using the adaptive observer, and an error is introduced via the history
stack since the state measurements recorded in the history stack are
corrupted by noise. The former can be addressed if a noise rejecting
observer such as a Kalman filter is used to generate the state-derivative
estimates. The latter can be addressed by the use of an inherently
noise-robust function approximation technique, e.g., a Gaussian process,
to approximate the system dynamics. An extension of the developed
parameter estimator that is provably robust to measurement noise is
a topic for future research. 

\bibliographystyle{IEEEtran}
\bibliography{encr,master,ncr}

\end{document}